\documentclass[12pt]{article}
\usepackage{latexsym,amssymb,amsfonts,amsmath}
\usepackage{mathrsfs}
\usepackage[dvips]{graphicx}
\newlength{\extraspace}
\newlength{\extraspaces}
\newcommand{\be}{\begin{equation}
\addtolength{\abovedisplayskip}{\extraspaces}
\addtolength{\belowdisplayskip}{\extraspaces}
\addtolength{\abovedisplayshortskip}{\extraspace}
\addtolength{\belowdisplayshortskip}{\extraspace}}
\newcommand{\ee}{\end{equation}}

\newcommand{\bea}{\begin{eqnarray}
\addtolength{\abovedisplayskip}{\extraspaces}
\addtolength{\belowdisplayskip}{\extraspaces}
\addtolength{\abovedisplayshortskip}{\extraspace}
\addtolength{\belowdisplayshortskip}{\extraspace}}
\newcommand{\eea}{\end{eqnarray}}

\DeclareMathOperator{\Tr}{Tr} 
\DeclareMathOperator{\diag}{diag} 

\newcommand{\Overline}[2][1]{%
 {}\mkern#1mu \overline{\mkern-#1mu #2 \mkern-#1mu}\mkern#1mu {}}

\newcommand{\rad}{\sqrt}
\newcommand{\de}{\partial}

\begin{document}

\addtolength{\baselineskip}{.8mm}

{\thispagestyle{empty}


\begin{center}
\vspace*{1.0cm}
{\Large\bf Study (using a chiral effective Lagrangian model)\\ of the scalar and pseudoscalar meson mass spectrum of QCD at finite temperature, above $T_c$}\\
\vspace*{1.0cm}
{\large
Enrico Meggiolaro\footnote{E-mail: enrico.meggiolaro@unipi.it}
}\\
\vspace*{0.5cm}{\normalsize
{Dipartimento di Fisica, Universit\`a di Pisa,
and INFN, Sezione di Pisa,\\
Largo Pontecorvo 3, I-56127 Pisa, Italy}}\\
\vspace*{2cm}{\large \bf Abstract}
\end{center}

\noindent
In this work, we analyze (using a chiral effective Lagrangian model) the scalar and pseudoscalar meson mass spectrum of QCD at finite temperature, above the chiral transition at $T_c$, in the realistic case with $N_f = 2 + 1$ light quark flavors (that is, with $m_{u,d} \to 0$ and $m_s \neq 0$), looking, in particular, for signatures of the breaking of the $U(1)$ axial symmetry above $T_c$.
A critical comparison with the corresponding results obtained (in a previous study) in the two limit cases $N_f = 2$ and $N_f = 3$ and with the available lattice results is performed, together with a discussion on the limits of validity of a description in terms of a chiral effective Lagrangian model.

}
\newpage

\section{Introduction}

Nowadays, lattice simulations \cite{HotQCD} provide a clear evidence of the fact that, at temperatures above a certain critical temperature $T_c \approx 150$ MeV, thermal fluctuations break up the chiral condensate $\langle \bar{q} q \rangle$, causing the complete restoration of the $SU(N_f)_L\otimes SU(N_f)_R$ chiral symmetry of QCD with $N_f$ light (massless) quarks ($N_f=2$ and $N_f=3$ being the physically relevant cases): this gives rise to a phase transition called ``chiral transition''.
Instead, the $U(1)$ axial symmetry is broken by a quantum anomaly and, because of the nonzero (even if decreasing) contribution to the anomaly provided by the instanton gas at high temperatures \cite{GPY1981}, it is expected to be always broken, also for $T>T_c$. (However, the real magnitude of its breaking and its possible \emph{effective} restoration at some temperature above $T_c$ are still important debated questions in hadronic physics: see Ref. \cite{CM2022} for a recent review on this subject.)

In this work, we analyze (using a chiral effective Lagrangian model) the scalar and pseudoscalar meson mass spectrum of QCD at finite temperature, above the chiral transition at $T_c$, in the realistic case with $N_f = 2 + 1$ light quark flavors (that is, with $m_{u,d} \to 0$ and $m_s \neq 0$), looking, in particular, for signatures of the breaking of the $U(1)$ axial symmetry above $T_c$.
A critical comparison with the corresponding results obtained in Ref. \cite{MM2013} in the two limit cases $N_f = 2$ and $N_f = 3$ is performed, together with a discussion on the limits of validity of a description in terms of a chiral effective Lagrangian model.


The effective Lagrangian model that we shall consider is well known: it was originally proposed in Ref. \cite{ELSM1} to study the chiral dynamics at $T=0$ (see also Refs. \cite{ELSM2,ELSM3}), and later used as an effective model to study the chiral-symmetry restoration at nonzero temperature \cite{PW1984,ELSMfiniteT_1,ELSMfiniteT_2}.
For brevity, following the notation already introduced in Refs. \cite{LM2018,EM2019}, we shall refer to it as the ``extended linear sigma ($EL_\sigma$) model''.\footnote{In Ref. \cite{MM2013} a different effective Lagrangian model was used, which, according to the notation already introduced in Refs. \cite{LM2018,EM2019}, is known as the ``interpolating model'', because, in a sense which is explained in those two references, it approximately \emph{interpolates}, when varying the temperature, between the effective Lagrangian model proposed by Witten, Di Vecchia, Veneziano, \emph{et al.} \cite{WDV1,WDV2,WDV3} at $T=0$ and the above-mentioned ``extended linear sigma model'' for $T>T_c$. Since in this paper we are interested in deriving the results for the mass spectrum only for temperatures above the transition, we have decided to use (for simplicity) the ``extended linear sigma model''.}
This model is described by the following Lagrangian:
\begin{equation}\label{ELSM Lagrangian}
\mathscr{L}_{(EL_\sigma)}(U,U^{\dagger}) = \frac{1}{2}\Tr [\partial_\mu U \partial^\mu U^{\dagger}] - V(U,U^{\dagger}) ,
\end{equation}
where
\begin{equation}\label{ELSM potential}
\begin{aligned}
V(U,U^\dagger) &= \frac{1}{4}\lambda_\pi^2 \Tr \left[(UU^\dagger - \rho_\pi \mathbf{I})^2\right] + \frac{1}{4}\lambda_\pi^{'2} \left[\Tr(UU^\dagger)\right]^2 \\ &- \frac{B_m}{2\sqrt{2}}\Tr \left[M U + M^\dagger U^\dagger\right] - \kappa \left[\det U + \det U^\dagger\right] .
\end{aligned}
\end{equation}
In this model, the mesonic effective fields are represented by a $N_f\times N_f$ complex matrix $U_{ij}$ which can be written in terms of the quark fields as $U_{ij}\sim \Overline[2]{q}_{jR}q_{iL}$, up to a multiplicative constant; moreover, $M = \diag (m_1, \ldots, m_{N_f})$ is the quark mass matrix.\\
Since under $U(N_f)_L\otimes U(N_f)_R$ chiral transformations the quark fields and the mesonic effective field $U$ transform as
\begin{equation}\label{trasfU}
U(N_f)_L\otimes U(N_f)_R:\quad q_{L,R}\rightarrow V_{L,R}q_{L,R} ~\Rightarrow~ U\rightarrow V_{L} U V_{R}^{\dagger} ,
\end{equation}
where $V_L$ and $V_R$ are arbitrary $N_f \times N_f$ unitary matrices, we have that the first term in the right-hand side of Eq. \eqref{ELSM Lagrangian} and the two first terms in the right-hand side of Eq. \eqref{ELSM potential} are invariant under the entire chiral group $U(N_f)_L\otimes U(N_f)_R$, while the last (\emph{anomalous}) interaction term, proportional to the parameter $\kappa$, in the right-hand side of Eq. \eqref{ELSM potential} [and so the entire effective Lagrangian \eqref{ELSM Lagrangian} in the \emph{chiral limit} $M=0$] is invariant under $SU(N_f)_L\otimes SU(N_f)_R \otimes U(1)_V$ but \emph{not} under a $U(1)$ axial transformation:
\begin{equation}\label{U1A}
U(1)_A:\quad q_{L,R}\rightarrow e^{\pm i\alpha}q_{L,R} ~\Rightarrow~
U \rightarrow e^{i2\alpha} U .
\end{equation}
Therefore, the term proportional to the parameter $\kappa$ in the potential \eqref{ELSM potential} describes (at the level of our effective Lagrangian model) the effects of the $U(1)$ axial anomaly.\\
For what concerns the other terms in the potential \eqref{ELSM potential}, we recall that the parameter $\rho_\pi$ is responsible for the fate of the $SU(N_f)_L \otimes SU(N_f)_R$ chiral symmetry, which, as we have already said, depends on the temperature $T$.
The effects of the temperature can be included in the model by considering the various parameters in Eq. \eqref{ELSM potential} as functions of the temperature: in particular, the parameter $\rho_\pi$ will be positive, and, correspondingly, the ``vacuum expectation value'' (vev), i.e., the thermal average, of $U$ will be different from zero in the chiral limit $M=0$, until the temperature reaches the chiral phase-transition temperature $T_c^{(N_f)}$ [$\rho_\pi(T<T_c^{(N_f)})>0$], above which it will be negative [$\rho_\pi(T>T_c^{(N_f)})<0$], and, correspondingly, the vev of $U$ will vanish in the chiral limit $M=0$.\footnote{We notice here that we have identified the temperature $T_{\rho_\pi}^{(N_f)}$ at which the parameter $\rho_\pi$ is equal to zero with the chiral phase-transition temperature $T_c^{(N_f)}$: this is surely correct for $N_f \ge 3$, but not in the special case $N_f=2$, where we have $T_{\rho_\pi}^{(2)}<T_c^{(2)}$ (see Refs. \cite{MM2013,EM2019} for a more detailed discussion). The special case $N_f=2+1$ will be discussed in the next section.}

\section{Mass spectrum for $T>T_c$ in the case $N_f=2+1$}

For simplicity, we shall consider the case of massless \emph{up} and \emph{down} quarks ($m_u = m_d = 0$) and a massive \emph{strange} quark:
\begin{equation}\label{M}
M =
\begin{pmatrix}
0 & 0 & 0 \\
0 & 0 & 0 \\ 
0 & 0 & m_s
\end{pmatrix} .
\end{equation}
Let us observe that, in this case, all the terms in the right-hand sides of Eqs. \eqref{ELSM Lagrangian} and \eqref{ELSM potential}, with the exception of the \emph{anomalous} interaction term, proportional to the parameter $\kappa$, are invariant under the (sub-)group of chiral transformations $U(2)_L^{(u,d)} \otimes U(2)_R^{(u,d)}$: $q_{L,R}\rightarrow V_{L,R}^{(u,d)}q_{L,R} ~\Rightarrow~ U\rightarrow V_{L}^{(u,d)} U V_{R}^{(u,d) \dagger}$, where $V_L^{(u,d)}$ and $V_R^{(u,d)}$ are arbitrary unitary matrices which only mix the \emph{up} and \emph{down} quark components, leaving the \emph{strange} quark component unaltered [see, for example, Eqs. \eqref{SU2A_ud} and \eqref{U1A_ud} below].
Instead, the (\emph{anomalous}) interaction term $\kappa \left[ \det U + \det U^\dagger \right]$ [and so the entire effective Lagrangian \eqref{ELSM Lagrangian}] is invariant under $SU(2)_L^{(u,d)}\otimes SU(2)_R^{(u,d)} \otimes U(1)_V^{(u,d)}$ but \emph{not} under $U(1)_A^{(u,d)}$.\\
We expect that the $SU(2)_L^{(u,d)} \otimes SU(2)_R^{(u,d)}$ chiral symmetry, while being spontaneously broken at $T=0$, will be restored above a certain critical temperature $T_c \equiv T_c^{(2)}$ [which does not necessarily coincide with $T_c^{(3)}$, the critical temperature in the case of $N_f=3$ massless flavors: their relation will be discussed below, immediately after Eq. \eqref{condition_Tc}].


For $T>T_c$, it is convenient to use for the matrix field $U$ the following {\it linear} parametrization, explicitly written in terms of the fields describing the relevant scalar and pseudoscalar mesonic excitations:
\begin{eqnarray}\label{Ulinear}
U &=& \frac{1}{\rad2}\sum_{a=1}^8 (h_a + i\pi_a)\tau_a + \frac{1}{\rad{3}}(h_0 + iS_\pi)\mathbf{I} \nonumber \\
&=& \frac{1}{\rad2}\sum_{a=1}^7 (h_a + i\pi_a)\tau_a + \frac{1}{\rad{2}}(\sigma_2 + i\eta_2)\mathbf{I}_2 + (\sigma_s + i\eta_s)\mathbf{I}_s ,
\end{eqnarray}
where $\mathbf{I}$ is the $3 \times 3$ identity matrix, $\tau_a$ ($a=1,\dots,8$) are the eight Gell-Mann matrices, with the usual normalization $\Tr(\tau_a\tau_b)=2\delta_{ab}$, and
\begin{equation}\label{matrices I_2 and I_s}
\mathbf{I}_2 \equiv
\begin{pmatrix}
1 & 0 & 0 \\
0 & 1 & 0 \\ 
0 & 0 & 0
\end{pmatrix} ,\qquad
\mathbf{I}_s \equiv
\begin{pmatrix}
0 & 0 & 0 \\
0 & 0 & 0 \\ 
0 & 0 & 1
\end{pmatrix} .
\end{equation}
That is, more explicitly, $U = H_S + i H_{PS}$, where $H_S$ and $H_{PS}$ are two Hermitian matrices defined respectively as
\begin{equation}\label{H_S}
H_S \equiv
\begin{pmatrix}
\frac{1}{\rad2} \delta^0 + \frac{1}{\rad2} \sigma_2 & \delta^+ & \kappa^+ \\
\delta^- & -\frac{1}{\rad2} \delta^0 + \frac{1}{\rad2} \sigma_2 & \kappa^0 \\ 
\kappa^- & \bar\kappa^0 & \sigma_s
\end{pmatrix}
\end{equation}
and
\begin{equation}\label{H_PS}
H_{PS} \equiv
\begin{pmatrix}
\frac{1}{\rad2} \pi^0 + \frac{1}{\rad2} \eta_2 & \pi^+ & K^+ \\
\pi^- & -\frac{1}{\rad2} \pi^0 + \frac{1}{\rad2} \eta_2 & K^0 \\ 
K^- & \bar{K}^0 & \eta_s
\end{pmatrix} ,
\end{equation}
where the fields in $H_S$ (linear combinations of $h_a$, $h_0$) are \emph{scalar} ($J^P = 0^+$) mesonic fields, while the fields in $H_{PS}$ (linear combinations of $\pi_a$, $S_\pi$) are \emph{pseudoscalar} ($J^P = 0^-$) mesonic fields.
In particular, the pseudoscalar fields $\pi^0 \equiv \pi_3$, $\pi^\pm \equiv (\pi_1 \mp i\pi_2)/\rad2$ correspond to the \emph{pions} and the pseudoscalar fields $K^\pm \equiv (\pi_4 \mp i\pi_5)/\rad2$, $K^0 \equiv (\pi_6 - i\pi_7)/\rad2$, and $\bar{K}^0 \equiv (\pi_6 + i\pi_7)/\rad2$ correspond to the \emph{kaons},
while the scalar fields $\delta^0 \equiv h_3$, $\delta^\pm \equiv (h_1 \mp ih_2)/\rad2$ correspond to the scalar partners of pions and the scalar fields $\kappa^\pm \equiv (h_4 \mp ih_5)/\rad2$, $\kappa^0 \equiv (h_6 - ih_7)/\rad2$, and $\bar{\kappa}^0 \equiv (h_6 + ih_7)/\rad2$ correspond to the scalar partners of the kaons.
Moreover, the two scalar fields $\sigma_2$ and $\sigma_s$ are the two following linear combinations of $h_8$ and $h_0$:
\begin{equation}\label{sigma_2,sigma_s}
\sigma_2 \equiv \frac{1}{\rad3} h_8 + \sqrt{\frac{2}{3}} h_0 ,\quad
\sigma_s \equiv -\sqrt{\frac{2}{3}} h_8 + \frac{1}{\rad3} h_0 ,
\end{equation}
and, correspondingly, the two pseudoscalar fields $\eta_2$ and $\eta_s$ are the two following linear combinations of $\pi_8$ and $S_\pi$:
\begin{equation}\label{eta_2,eta_s}
\eta_2 \equiv \frac{1}{\rad3} \pi_8 + \sqrt{\frac{2}{3}} S_\pi ,\quad
\eta_s \equiv -\sqrt{\frac{2}{3}} \pi_8 + \frac{1}{\rad3} S_\pi .
\end{equation}
Let us observe that
$h_0,~S_\pi \sim \frac{1}{\rad3} (u\bar{u} + d\bar{d} + s\bar{s})$
are scalar and pseudoscalar $SU(3)$ singlet fields,
and $h_8,~\pi_8 \sim \frac{1}{\sqrt{6}} (u\bar{u} + d\bar{d} - 2 s\bar{s})$
are scalar and pseudoscalar $SU(3)$ octet fields,
while $\sigma_2$, $\eta_2 \sim \frac{1}{\rad2} (u\bar{u} + d\bar{d})$
are scalar and pseudoscalar $SU(2)$ singlet fields
and $\sigma_s$, $\eta_s \sim s\bar{s}$.\\
We also recall that, under $SU(2)_A^{(u,d)}$ and $U(1)_A^{(u,d)}$ transformations, defined as
\begin{equation}\label{SU2A_ud}
SU(2)_A^{(u,d)}:\quad V_L = V_R^\dagger =
\begin{pmatrix}
A_{11} & A_{12} & 0 \\
A_{21} & A_{22} & 0 \\ 
0 & 0 & 1
\end{pmatrix} ,\quad {\rm with}:\quad \tilde{A} \equiv
\begin{pmatrix}
A_{11} & A_{12} \\
A_{21} & A_{22}
\end{pmatrix} \in SU(2) ,
\end{equation}
and
\begin{equation}\label{U1A_ud}
U(1)_A^{(u,d)}:\quad V_L = V_R^\dagger =
\begin{pmatrix}
e^{i\alpha} & 0 & 0 \\
0 & e^{i\alpha} & 0 \\ 
0 & 0 & 1
\end{pmatrix} ,
\end{equation}
the meson channels $\sigma_2$, $\eta_2$, $\vec\pi$, and $\vec\delta$ are mixed as follows:
\begin{equation}
\begin{matrix}
\sigma_2 & \stackrel{U(1)_A^{(u,d)}}{\longleftrightarrow} & \eta_2 \\
SU(2)_A^{(u,d)} \updownarrow & & \updownarrow SU(2)_A^{(u,d)} \\
\vec{\pi} & \stackrel{U(1)_A^{(u,d)}}{\longleftrightarrow} & \vec{\delta}
\end{matrix}
\end{equation}
The restoration of the $SU(2)_L^{(u,d)} \otimes SU(2)_R^{(u,d)}$ chiral symmetry implies that the $\sigma_2$ and $\vec\pi$ channels become degenerate, with identical masses $M_{\sigma_2} = M_\pi$, and the same happens also for the channels $\eta_2$ and $\vec\delta$.
Instead, an \emph{effective restoration} of the $U(1)^{(u,d)}$ axial symmetry should imply that $\sigma_2$ becomes degenerate with $\eta_2$, and $\vec\pi$ becomes degenerate with $\vec\delta$.
(Clearly, if both chiral symmetries were restored, then all $\sigma_2$, $\vec\pi$, $\eta_2$, and $\vec\delta$ masses should become the same.)


For $T>T_c$, the value $\Overline[2]{U}$ for which the potential $V$ is minimum (that is, in our mean-field approach, the vev of $U$), is given by:
\begin{equation}\label{U-bar}
\Overline[2]{U} =
\begin{pmatrix}
0 & 0 & 0 \\
0 & 0 & 0 \\ 
0 & 0 & \bar\sigma_s
\end{pmatrix} ,
\end{equation}
where $\bar\sigma_s$ (the vev of the hermitian field $\sigma_s$) must be a \emph{real} solution of the following equation (\emph{stationary-point condition}):
\begin{equation}\label{sigma_s-bar}
\left. \frac{\partial V}{\partial \sigma_s}\right|_{SP} = (\lambda_\pi^2 + \lambda_\pi^{\prime 2}) \bar\sigma_s^3 - \lambda_\pi^2 \rho_\pi \bar\sigma_s - \frac{1}{\rad 2} B_m m_s = 0 .
\end{equation}
All the other first derivatives of the potential with respect to the various scalar and pseudoscalar meson fields come out to be trivially zero when evaluated at the \emph{stationary point} ($SP$) given by Eq. \eqref{U-bar}.\\
The quantity $\bar\sigma_s$ turns out to be proportional to the strange-quark chiral condensate $\langle \bar{s} s \rangle$, being
\begin{equation}\label{chiralcondensate}
\langle \bar{s} s \rangle = \frac{\de \Overline[2]{V}}{\de m_s} = -\frac{1}{\rad2} B_m \bar\sigma_s ,
\end{equation}
where $\Overline[2]{V} = -\frac{B_m}{2\rad 2} \Tr[M(\Overline[2]{U}+\Overline[2]{U}^{\dagger})] + \dots = -\frac{1}{\rad 2}B_m m_s \bar\sigma_s + \dots$ is the vacuum expectation value of the potential of the effective Lagrangian.

The squared masses of the various scalar and pseudoscalar mesonic excitations, corresponding to the second derivatives of the potential with respect to the fields [evaluated at the \emph{stationary point} \eqref{U-bar}], come out to be:
\begin{equation}\label{masses}
\begin{aligned}
M_{\sigma_2}^2 = M_\pi^2 &= -\lambda_\pi^2 \tilde\rho_\pi - 2\tilde\kappa ,\\
M_{\eta_2}^2 = M_\delta^2 &= -\lambda_\pi^2 \tilde\rho_\pi + 2\tilde\kappa ,\\
M_K^2 = M_\kappa^2 = M_{\eta_s}^2 &= -\lambda_\pi^2 \tilde\rho_\pi + \lambda_\pi^2 \bar\sigma_s^2 ,\\
M_{\sigma_s}^2 &= -\lambda_\pi^2 \tilde\rho_\pi + 3\lambda_\pi^2 \bar\sigma_s^2 + 2\lambda_\pi^{\prime 2} \bar\sigma_s^2 ,
\end{aligned}
\end{equation}
where
$M_\pi \equiv M_{\pi_1} = M_{\pi_2} = M_{\pi_3}$,
$M_\delta \equiv M_{h_1} = M_{h_2} = M_{h_3}$,
$M_K \equiv M_{\pi_4} = M_{\pi_5} = M_{\pi_6} = M_{\pi_7}$,
$M_\kappa \equiv M_{h_4} = M_{h_5} = M_{h_6} = M_{h_7}$,
and, moreover,
\begin{equation}\label{rho-tilde,kappa-tilde}
\tilde\rho_\pi \equiv \rho_\pi - \frac{\lambda_\pi^{\prime 2}}{\lambda_\pi^2} \bar\sigma_s^2 ,\quad \tilde\kappa \equiv \kappa \bar\sigma_s .
\end{equation}
Of course, the above-reported solution is acceptable only if it really corresponds to a \emph{minimum} of the potential, i.e., only if all the squared masses in Eq. \eqref{masses} (corresponding to the second derivatives of the potential) are positive (or, at least, nonnegative). This implies that the following necessary (but not sufficient) condition must be satisfied:
\begin{equation}\label{condition_rho-tilde}
\tilde\rho_\pi \equiv \rho_\pi - \frac{\lambda_\pi^{\prime 2}}{\lambda_\pi^2} \bar\sigma_s^2 \le 0 ,
\end{equation}
since, otherwise, the sum $M_\pi^2 + M_\delta^2 = M_{\sigma_2}^2 + M_{\eta_2}^2$ would be negative\dots\\
For example, in the (ideal) limit case $m_s=0$, recalling that $\rho_\pi(T)>0$ for $T<T_c^{(3)}$ and $\rho_\pi(T)<0$ for $T>T_c^{(3)}$ (being $T_c^{(3)}$ the critical temperature in the case of $N_f=3$ massless flavors), one immediately sees that there is only one real solution of Eq. \eqref{sigma_s-bar} corresponding to a minimum of the potential for $T\ge T_c^{(3)}$ and it is given by $\bar\sigma_s=0$. [Instead, there is no acceptable solution of the type \eqref{U-bar} for $T<T_c^{(3)}$, since in this case one would obtain that, for each real solution of Eq. \eqref{sigma_s-bar} with $m_s=0$, $\tilde\rho_\pi \equiv \rho_\pi - \frac{\lambda_\pi^{\prime 2}}{\lambda_\pi^2} \bar\sigma_s^2 > 0$, i.e., the necessary condition \eqref{condition_rho-tilde} for a solution corresponding to a miminum of the potential is \emph{not} satisfied\dots]
In this case ($m_s=0$, $\bar\sigma_s=0$, i.e., $\Overline[2]{U}=0$, for $T\ge T_c^{(3)}$) one obtains a solution correponding (at least for what concerns the mass spectrum) to a restoration of the entire $U(3) \otimes U(3)$ chiral symmetry, with all the various scalar and pseudoscalars mesonic excitations being degenerate, with a common value of the squared mass given by
\begin{equation}\label{masses_m_s=0}
M_{\sigma_2}^2 = M_\pi^2 = M_{\eta_2}^2 = M_\delta^2 = M_K^2 = M_\kappa^2 = M_{\eta_s}^2 = M_{\sigma_s}^2 = -\lambda_\pi^2 \tilde\rho_\pi = -\lambda_\pi^2 \rho_\pi .
\end{equation}
The situation in the realistic case $m_s \neq 0$ is, of course, more complicated. First of all, let us observe that Eq. \eqref{sigma_s-bar} always admits \emph{at least} one real and positive solution: in fact, denoting (for brevity) the function of $\bar\sigma_s$ in the left-hand side of Eq. \eqref{sigma_s-bar} with $F(\bar\sigma_s)$, one sees that $F(0) = -\frac{1}{\rad 2} B_m m_s < 0$, while $\displaystyle\lim_{x \to +\infty} F(x) = +\infty$.\\
Moreover, a real solution of Eq. \eqref{sigma_s-bar} \emph{must} be positive, provided that the necessary condition \eqref{condition_rho-tilde} (for a solution corresponding to a minimum of the potential) is satisfied. In fact, one immediately sees that $F(x) = \lambda_\pi^2 x^3 - \tilde\rho_\pi(x) x - \frac{1}{\rad 2} B_m m_s \le -\frac{1}{\rad 2} B_m m_s < 0$, if $x<0$ and $\tilde\rho_\pi(x) \equiv \rho_\pi - \frac{\lambda_\pi^{\prime 2}}{\lambda_\pi^2} x^2 \le 0$.
In other words, since the solution of Eq. \eqref{sigma_s-bar} which corresponds to the minimum of the potential must necessarily satisfy the condition \eqref{condition_rho-tilde}, it must be real and positive: $\bar\sigma_s > 0$.\\
This solution, that is the expression \eqref{U-bar} of $\Overline[2]{U}$ with this value of $\bar\sigma_s$, describes (in the case in which $\kappa>0$) the restoration of the $SU(2)_L^{(u,d)} \otimes SU(2)_R^{(u,d)}$ chiral symmetry above a certain critical temperature $T_c$ defined by the condition:
\begin{equation}\label{condition_Tc}
M_\pi^2(T_c) = -\lambda_\pi^2 \rho_\pi(T_c) + \lambda_\pi^{\prime 2} \bar\sigma_s^2(T_c) - 2\kappa \bar\sigma_s(T_c) = 0 ,
\end{equation}
and, of course, 
$M_\pi^2 = -\lambda_\pi^2 \tilde\rho_\pi - 2\tilde\kappa = -\lambda_\pi^2 \rho_\pi + \lambda_\pi^{\prime 2} \bar\sigma_s^2 - 2\kappa \bar\sigma_s \ge 0$ for $T\ge T_c$.
From the condition \eqref{condition_Tc} one sees that the critical temperature $T_c$ would coincide with $T_c^{(3)}$ in the particular limit case in which $\lambda_\pi^\prime = \kappa = 0$ (while keeping $m_s\neq 0$; of course, as we have already seen above, they would also coincide in the limit case $m_s=0$). Instead, in the general case (with nonzero $\lambda_\pi^\prime$ and $\kappa$) the relation between $T_c$ and $T_c^{(3)}$ is more complicated, resulting from the competition between two opposite effects from a nonzero value of the parameter $\lambda_\pi^\prime$ and from a nonzero value of the parameter $\kappa$: more precisely, it is easy to see from Eq. \eqref{condition_Tc} that a nonzero value of the parameter $\lambda_\pi^\prime$ has the effect of \emph{decreasing} the value of the critical temperature $T_c$ with respect of $T_c^{(3)}$, while a nonzero value of the parameter $\kappa$ has the opposite effect of \emph{increasing} the critical temperature $T_c$ with respect of $T_c^{(3)}$.\\
Moreover, assuming for simplicity that the parameters $\lambda_\pi$, $\lambda_\pi^\prime$, and $B_m$ are approximately independent of $T$, at least in the vicinity of the critical temperature, and assuming for the parameter $\rho_\pi(T)$ a standard behavior of the form
\begin{equation}\label{rho_pi}
\rho_\pi(T) \simeq -a (T-T_c^{(3)}) ,\quad {\rm with} \quad a>0 ,
\end{equation}
one finds, deriving the two sides of Eq. \eqref{sigma_s-bar} with respect of the temperature $T$, that the solution $\bar\sigma_s(T)$ of such equation is a decreasing function of $T$, being:
\begin{equation}\label{derivative_sigma_s-bar}
\frac{d\bar\sigma_s}{dT} \simeq -\frac{a \lambda_\pi^2 \bar\sigma_s}{3(\lambda_\pi^2 + \lambda_\pi^{\prime 2}) \bar\sigma_s^2 - \lambda_\pi^2 \rho_\pi}
= -\frac{a \lambda_\pi^2 \bar\sigma_s}{M_{\sigma_s}^2} < 0 .
\end{equation}
Using this result and making also the plausible assumption (confirmed by instanton and lattice calculations) that the effects of the anomaly decrease with the temperature, i.e., that ${d\kappa}/{dT} < 0$, we can also derive the (approximate) temperature dependence of the parameters $\tilde\rho_\pi$ and $\tilde\kappa$, defined in Eq. \eqref{rho-tilde,kappa-tilde}. Concerning $\tilde\kappa$, we find that (being also $\kappa > 0$ and $\bar\sigma_s > 0$)
\begin{equation}\label{derivative_kappa-tilde}
\frac{d\tilde\kappa}{dT} = \frac{d\kappa}{dT} \bar\sigma_s + \kappa \frac{d\bar\sigma_s}{dT} < 0 .
\end{equation}
And, concerning $\tilde\rho_\pi$, we find that:
\begin{equation}\label{derivative_rho-tilde}
\frac{d\tilde\rho_\pi}{dT} = \frac{d\rho_\pi}{dT} - \frac{\lambda_\pi^{\prime 2}}{\lambda_\pi^2} 2\bar\sigma_s \frac{d\bar\sigma_s}{dT} \simeq
-a \frac{3\lambda_\pi^2 \bar\sigma_s^2 + \lambda_\pi^{\prime 2} \bar\sigma_s^2 - \lambda_\pi^2 \rho_\pi}{M_{\sigma_s}^2} < 0 ,
\end{equation}
where we have used Eq. \eqref{masses} to prove that the numerator in the fraction in the last equation is positive, being
$M_{\eta_s}^2 < 3\lambda_\pi^2 \bar\sigma_s^2 + \lambda_\pi^{\prime 2} \bar\sigma_s^2 - \lambda_\pi^2 \rho_\pi < M_{\sigma_s}^2$ (and the fraction is therefore a positive number smaller than 1).\\
Finally, using the results \eqref{derivative_kappa-tilde} and \eqref{derivative_rho-tilde}, we derive that the squared mass of the pion is an \emph{increasing} function of the temperature $T$:
\begin{equation}\label{derivative_pion-mass}
\frac{d M_\pi^2}{dT} \simeq -\lambda_\pi^2 \frac{d\tilde\rho_\pi}{dT} - 2\frac{d\tilde\kappa}{dT} > 0 .
\end{equation}

\section{Conclusions: summary and analysis of the results}

Eq. \eqref{masses} represents the main result obtained in this paper and we want now to make some comments on it.
First of all, let us observe that the anomalous term (proportional to the parameter $\kappa$) in the effective Lagrangian, influences the mass spectrum only of the \emph{nonstrange} (NS) mesonic excitations, where two $(\frac{1}{2},\frac{1}{2})$ chiral multiplets appear, namely $(\sigma_2,\vec\pi)$ and $(\eta_2,\vec\delta)$, signalling the restoration of the $SU(2)_L^{(u,d)} \otimes SU(2)_R^{(u,d)}$ chiral symmetry, while the squared masses of the nonstrange mesonic excitations belonging to a same $U(1)_A^{(u,d)}$ chiral multiplet, such as $(\sigma_2,\eta_2)$ and $(\vec\pi,\vec\delta)$, are splitted by the quantity:
\begin{equation}\label{mass-split_1}
\Delta M_{U(1)_A^{(u,d)}}^2 \equiv M_{\eta_2}^2 - M_{\sigma_2}^2 = M_\delta^2 - M_\pi^2 = 4\tilde\kappa = 4\kappa\bar\sigma_s .
\end{equation}
We have thus found, concerning the nonstrange mesonic excitations, the same results that were found in Ref. \cite{MM2013} using a chiral effective Lagrangian with $N_f=2$ massless flavors: the quantities $\tilde\rho_\pi$ and $\tilde\kappa$, defined in Eq. \eqref{rho-tilde,kappa-tilde}, are indeed the quantities that replace, in the ``reduced'' $N_f=2$ effective Lagrangian, the corresponding quantities $\rho_\pi$ and $\kappa$ in the $N_f=3$ effective Lagrangian\dots\footnote{In order to perform the \emph{formal} limit $m_s\to\infty$, leading to the complete \emph{decoupling} of the \emph{strange} quark, i.e., of all the scalar and pseudoscalar mesons including the \emph{strange} quark, one needs to \emph{renormalize} the parameters $\rho_\pi$ and $\kappa$ in order to keep the quantities $\tilde\rho_\pi$ and $\tilde\kappa$ \emph{fixed}
($\rho_\pi = \tilde\rho_\pi + \frac{\lambda_\pi^{\prime 2}}{\lambda_\pi^2} \bar\sigma_s^2$, $\kappa = \frac{\tilde\kappa}{\bar\sigma_s}$).
Observing that, by virtue of the definition of $\tilde\rho_\pi$, Eq. \eqref{sigma_s-bar} can be rewritten as
$\lambda_\pi^2 \bar\sigma_s^3 - \lambda_\pi^2 \tilde\rho_\pi \bar\sigma_s - \frac{1}{\rad 2} B_m m_s = 0$, one finds the following asymptotic behavior of the solution $\bar\sigma_s$ in the limit $m_s\to\infty$:
\begin{equation}\label{sigma_s-bar_asympt}
\bar\sigma_s \mathop\simeq_{m_s\to\infty} \left( \frac{B_m m_s}{\rad2 \lambda_\pi^2} \right)^{1/3} .
\end{equation}
Therefore:
\begin{equation}\label{rho-pi_kappa_renormalization}
\rho_\pi = \tilde\rho_\pi + \frac{\lambda_\pi^{\prime 2}}{\lambda_\pi^2} \bar\sigma_s^2 \mathop\simeq_{m_s\to\infty}
\tilde\rho_\pi + \frac{\lambda_\pi^{\prime 2}}{\lambda_\pi^2} \left( \frac{B_m m_s}{\rad2 \lambda_\pi^2} \right)^{2/3} ,\quad {\rm and} \quad
\kappa = \frac{\tilde\kappa}{\bar\sigma_s} \mathop\simeq_{m_s\to\infty} \tilde\kappa \left( \frac{\rad2 \lambda_\pi^2}{B_m m_s} \right)^{1/3}.
\end{equation}
By virtue of this, the squared masses of the mesons $K$, $\kappa$, $\eta_s$, and $\sigma_s$ (containing the \emph{strange} quark) tend to infinity (like $\bar\sigma_s^2$) when $m_s\to\infty$, while the squared masses of the nonstrange mesons $(\sigma_2, \vec\pi)$ and $(\eta_2, \vec\delta)$ are kept \emph{fixed} to the values $-\lambda_\pi^2 \tilde\rho_\pi - 2\tilde\kappa$ and $-\lambda_\pi^2 \tilde\rho_\pi + 2\tilde\kappa$ respectively.}

Several lattice computations of the (\emph{screening}) meson masses at finite temperature (for the case $N_f=2$ and also for the more realistic case $N_f=2+1$), exist in the literature, but the results achieved so far are not yet conclusive.
Most of the studies \cite{lat1997,lat1998,lat1999,lat2000,lat2000bis,lat2011,lat2012,lat2014,Ding2021} (using \emph{staggered quarks} or \emph{domain-wall quarks} on the lattice) find that the $U(1)_A^{(u,d)}$-breaking difference \eqref{mass-split_1} is still sensibly nonzero above the chiral transition, but some others \cite{Cossu2013,Aoki2012} (using the so-called \emph{overlap quarks} on the lattice) find that this quantity vanishes immediately above the transition, so indicating an \emph{effective} restoration of the $U(1)_A^{(u,d)}$ symmetry already at $T_c$, at least at the level of the mass spectrum of the meson channels.

In addition, looking also at the other mesonic excitations containing the \emph{strange} quark flavor, we want to emphasize that the $SU(2)_L^{(u,d)} \otimes SU(2)_R^{(u,d)}$ chiral restoration is also manifest in the degeracy of the (pseudoscalar) \emph{kaons} $K^\pm$ and $K^0$, $\bar{K}^0$ with their scalar partners $\kappa^\pm$ and $\kappa^0$, $\bar\kappa^0$. This $K$--$\kappa$ degeneracy was also derived in Ref. \cite{GR2018} in a completely different way, analyzing a particular set of QCD Ward Identities.\\
More in general, the following hierarchy in the mass spectrum appears from Eq. \eqref{masses}:
\begin{equation}\label{hierarchy}
\Overline[2]{M}_{NS}^2 \equiv \frac{M_\pi^2 + M_\delta^2}{2} = \frac{M_{\sigma_2}^2 + M_{\eta_2}^2}{2} < M_K^2 = M_\kappa^2 = M_{\eta_s}^2 < M_{\sigma_s}^2 ,
\end{equation}
with [in addition to the mass split already mentioned in Eq. \eqref{mass-split_1}]
\begin{equation}\label{mass-split_2}
M_{\sigma_s}^2 - M_{\eta_s}^2 > 2 (M_{\eta_s}^2 - \Overline[2]{M}_{NS}^2) .
\end{equation}
The mass spectrum of mesonic excitations containing both the light (\emph{up} and \emph{down}) and the \emph{strange} quark flavors was also studied in Ref. \cite{lat2011} by means of lattice simulations: the results are approximately consistent with the hierarchy shown in Eq. \eqref{hierarchy}, even if the degeracy between $M_K$ and $M_\kappa$ is manifest only for temperatures larger than 1.3 times the chiral crossover temperature. This could be due to the fact that the lattice computations in Ref. \cite{lat2011} were not performed approaching the \emph{chiral limit} ($m_{u,d} \to 0$), but along a so-called \emph{line of constant physics}, defined by a zero-temperature pion mass of about 220 MeV and a zero-temperature kaon mass of about 500 MeV. So, further lattice computations, exploring the mesonic mass spectrum also approaching the chiral limit, would be surely welcome in order to check the validity of our result \eqref{hierarchy}.
(And, on the other side, also an analytical estimate, obtained with the same techniques adopted in the present paper, of the effects on the mesonic mass spectrum of a \emph{small} but nonzero value of $m_{u,d}$ would be extremely useful, since it could be directly compared with the above-mentioned lattice results: this issue will be addressed in a future work.)

In the (ideal) limit case $m_s=0$, being $\bar\sigma_s=0$, all the above-reported mass splits vanish and all the various scalar and pseudoscalars mesonic excitations become degenerate, with a common value of the squared mass given by Eq. \eqref{masses_m_s=0}, in agreement with the results already found in Ref. \cite{MM2013}, using a chiral effective Lagrangian with $N_f\ge 3$ massless flavors.
In this case, the anomalous term (proportional to the parameter $\kappa$) in the effective Lagrangian has no effect on the mass spectrum, which appears as if the entire $U(3) \otimes U(3)$ chiral symmetry were restored\dots\\
This situation might in principle be approximately realized also in the realistic case with $m_s\neq 0$ for those temperatures for which $\bar\sigma_s$ is ``sufficiently small''. More precisely, one sees from Eq. \eqref{sigma_s-bar} that, whenever $\bar\sigma_s \ll \sqrt{|\rho_\pi|}$, then the solution of the equation is given approximately by
\begin{equation}\label{sigma_s-bar_approx}
\bar\sigma_s \simeq -\frac{B_m m_s}{\rad2 \lambda_\pi^2 \rho_\pi} ,
\end{equation}
recalling also that now the necessary condition \eqref{condition_rho-tilde} (for a solution corresponding to a miminum of the potential) reduces to $\rho_\pi < 0$, so guaranteeing the positivity of $\bar\sigma_s$.
The expression \eqref{sigma_s-bar_approx} is consistent with the ``smallness'' condition $\bar\sigma_s \ll \sqrt{|\rho_\pi|}$ provided that
\begin{equation}\label{m_s-condition}
m_s \ll \frac{\rad2 \lambda_\pi^2 |\rho_\pi|^{3/2}}{B_m} = \frac{\rad2 \lambda_\pi^2 (a|t|)^{3/2}}{B_m} ,
\end{equation}
having used the (approximate) expression \eqref{rho_pi} for $\rho_\pi(T)$, with $t \equiv T-T_c^{(3)}$. This is nothing but the well-known ``weak external field condition'' (considering the mass term in the effective Lagrangian as an ``external field'') in the Landau theory of critical phenomena \cite{LL1980}.
This condition is of course not realized in the vicinity of (i.e., immediately above) the transition [the condition \eqref{condition_Tc} defining $T_c$ is not consistent with the ``smallness'' condition $\bar\sigma_s \ll \sqrt{|\rho_\pi|}$\dots]: in this case, the situation concerning the mass spectrum is surely closer to the one realized in the theory with $N_f=2$ massless flavors.
In principle, the condition \eqref{m_s-condition} might be realized for large enough temperatures, much above the transition (so leading to a situation concerning the mass spectrum closer to the one realized in the theory with $N_f=3$ massless flavors): however, taking into account the quite large value of the \emph{strange} quark mass ($m_s \sim 100$ MeV), it might also well be that these temperatures are too much above the transition, well inside the quark-gluon plasma phase, where a simple description in terms of a chiral effective Lagrangian, like the one that we have adopted in this paper, is no more applicable.\\
Of course, only a more \emph{quantitative} approach to the problem, implying a numerical estimate of the various parameters appearing in the chiral effective Lagrangian around the transition, will allow to shed light on these questions, as well as on the above-mentioned comparison with the lattice results: future works are therefore expected along this line.

\newpage

\renewcommand{\Large}{\large}

\end{document}